\documentclass[a4paper]{VisionStyle}
\usepackage{epsfig}

\begin{document}

\title{Studying the Cosmic X-ray Background with XMM-Newton.}

\author{A.\,De Luca\inst{1,2} \and S.\,Molendi\inst{1}} 

\institute{  Istituto di Astrofisica Spaziale e Fisica Cosmica, Via Bassini 15, I-20133 Milano, Italy
\and
  Universit\`a di Milano Bicocca, Dipartimento di Fisica, Piazza della Scienza 3, I-20126 Milano, Italy
}

\maketitle 

\begin{abstract}
We present a work in progress aimed at measuring the spectrum of the Cosmic 
X-ray Background (CXB) with the EPIC detectors onboard XMM-Newton. Our study 
includes a detailed characterization of the EPIC non X-ray background, which is
crucial in making a robust measurement of the spectrum of CXB. We present 
preliminary results, based on the analysis of a set of Commissioning and 
Performance Verification high galactic latitude observations.

\end{abstract}

\section{Introduction}
\label{adeluca-F_sec:intro}

The discovery of a diffuse background radiation in the X-ray sky dates back
to the birth of X-ray astronomy: the first evidence was obtained by
\cite*{adeluca-F:gi62} during the same rocket experiment which led to the
discovery of Sco X-1, the first extra-solar X-ray source.
Later observations have demonstrated that the bulk of Cosmic X-ray 
Background (CXB)
above energies of $\approx$2 keV is of extragalactic origin, due to sources
below the detection threshold.
The first wide band measures of the CXB were 
made by HEAO$-$1 (1977): the CXB spectrum in the 2$\div$10 keV range was 
well described by a simple power law with photon index $\approx$ 1.4 
(\cite{adeluca-F:ma80}).
More recently, several investigations (ASCA GIS$/$SIS: \cite{adeluca-F:mi98}, 
\cite{adeluca-F:ge95}; ROSAT PSPC: \cite{adeluca-F:ge96}; SAX LECS/MECS: 
\cite{adeluca-F:ve99}) have confirmed  the spectral shape but have shown 
differences  of order 30\% in the normalization. \cite*{adeluca-F:ba00} 
showed that cosmic variance cannot account for the differences among the 
previous measures of the CXB intensity; such an uncertainty seriously 
affects the modelling and  
the interpretation of the CXB (see e.g. \cite{adeluca-F:po00}).
A new, reliable measure of the CXB is thus required to improve 
the overall understanding of its nature. \\
The XMM-EPIC instruments (\cite{adeluca-F:tu01}; \cite{adeluca-F:s01}) have
appropriate characteristics to study extended sources with low surface
brighteness, offering an unprecedented collecting area ($\approx 2000$ 
cm$^2$) and good spectral resolution (2\% @ 6 keV) over a broad energy 
range (0.2$\div$12 keV) and a wide field of view ($\approx$ 15 arcmin radius).
However these cameras suffer a rather high instrumental background (Non X-ray 
Background, NXB). A correct characterization and subtraction of the NXB 
component is thus the crucial step in order to study the lowest surface 
brighteness source of the sky. \\
In this work we deal only with the EPIC {\em MOS} cameras; the {\em pn} camera,
having different characteristics, will require a different approach.

\section{Characterization of Non X-ray Background}
\label{adeluca-F_sec:nxb}

\begin{figure}[ht]
  \begin{center}
    \epsfig{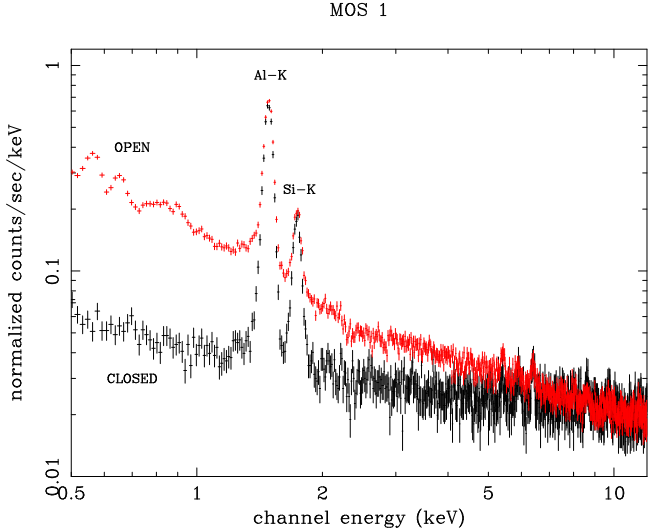}
  \end{center}
\caption{MOS1 quiescent NXB spectrum obtained from $\approx$100 ks of
closed observations. For a comparison, the spectrum extracted from $\approx$
 200 ks of high galactic latitude observations of blank fields 
is also plotted (labelled OPEN).
The closed spectrum is 
characterized by a flat continuum with bright fluorescence emission lines.
The open spectrum is the sum of CXB and quiescent NXB.}  
\label{adeluca-F_fig:spec1}
\end{figure}

The EPIC NXB can be divided into two parts: an electronic noise component, 
which is important only at the lower energies (below $\approx$0.3 keV), and  a 
particle-induced component which dominates above 0.3 keV 
and is due to the interaction of particles in the orbital environment with the 
detectors and the structures that surround them. 
The particle-induced NXB is the sum of two different components:
\begin{enumerate}
\item a {\em flaring} component. Clouds of low-energy ($\approx$ 100 keV) 
particles (believed to be protons) in the magnetosphere can be focused 
by the telescope mirrors, reaching the detectors. These unpredictable episodes
cause an up to 100 times (or even more) increase of the quiescent background 
rate. Data 
collected during these intervals are almost unusable, 
especially for the study of extended sources, and
must be rejected with Good Time Interval (GTI) filtering.
\item a {\em quiescent} component. It is mostly due to the interaction of high
energy (E $\geq$ a few MeV) particles with the detectors and the surrounding 
structures. To characterize the quiescent NXB we have analyzed a set of 
observations performed with the filter wheel in closed position:
\begin{itemize}
\item the temporal behaviour is stable within an observation time scale; we
have hints for a secular decrease in the period covered by the dataset (orbits
 20$\div$84);
\item the spectrum is characterized by a flat continuum, with several 
fluorescence emission lines from materials in the detectors or the 
surrounding structures (see Fig. \ref{adeluca-F_fig:spec1});
\item the spatial distribution has a radially flat profile (within 10 \%), but
Al-K and Si-K line emission are highly anisotropic due to an illumination 
effect.
\end{itemize}
\end{enumerate}
Further closed observation are now being collected to improve the quiescent
NXB characterization.

\section{Subtraction of Non X-ray Background}
\label{adeluca-F_sec:sub}
\begin{figure}[ht]
  \begin{center}
    \epsfig{file=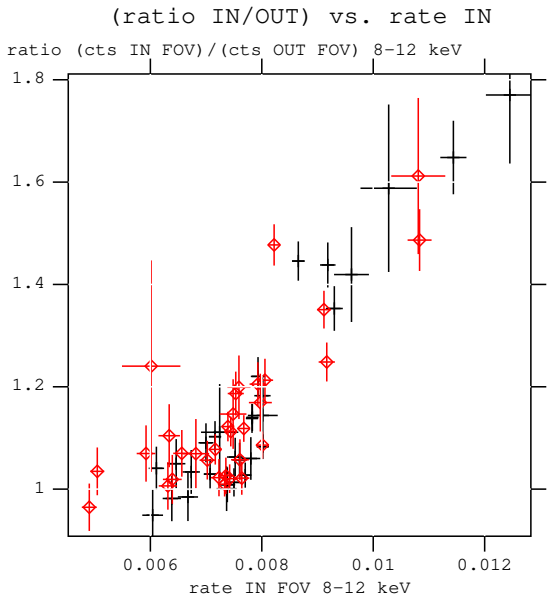, width=8cm}
  \end{center}
\caption{Plot of the ratio (cts IN FOV)/(cts OUT FOV) vs. count rate IN FOV. 
Each point represents an observation: MOS1 black crosses, MOS2 red diamonds.}  
\label{adeluca-F_fig:ratio}
\end{figure}

A standard recipe to remove the NXB recovering the ``pure'' CXB spectrum
could be sketched as follows:
\begin{enumerate}
\item standard processing and event reconstruction;
\item rejection of hot pixels and bad columns;
\item GTI filtering (removing the flaring NXB component);
\item extraction of the spectrum from a selected area;
\item subtraction of quiescent NXB spectrum.
\end{enumerate}
This algorithm is indeed quite dangerous, since GTI filtering possibly leaves 
a low level Soft Proton (SP) NXB component unrejected; such a component could 
affect the CXB spectrum determination. In order to avoid this, we developed
a simple diagnostic. 
\\ Since SP are focused by the mirrors, we have studied
the correlation between counts extracted from the regions of the detectors which
are inside (IN FOV) and outside (OUT FOV) the field of view of the telescopes.
We applied to the OUT FOV event lists the same GTI computed for the 
corresponding IN FOV data. For each observation we extracted the counts 
(PATTERN 0 $\div$ 4) in the $8 \div 12$ keV band (where the CXB is negligible wrt.
 the NXB) from appropriate regions: in the IN FOV region we rejected CCD 1, 
where possibly CXB is not negligible, and we applied geometric masks to screen 
the brightest sources; in both the IN FOV and the OUT FOV regions we further 
applied geometric masks to reject the region spatially coincident with the 
underlying CCD 1 readout node (Au line emission). 
\begin{figure}[ht]
  \begin{center}
    \epsfig{file=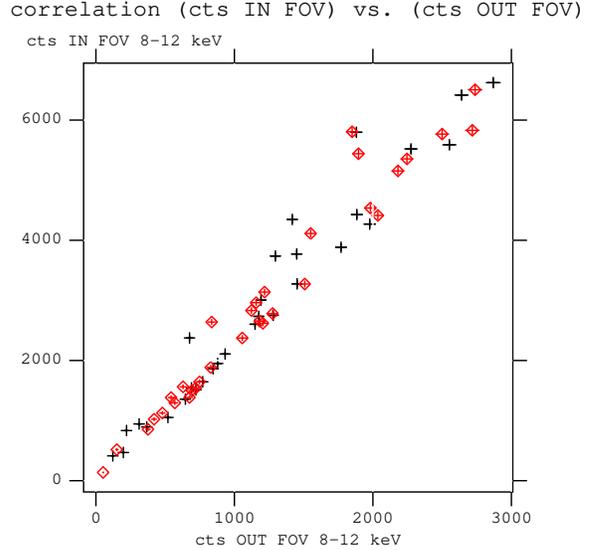, width=9cm}
  \end{center}
\caption{Plot of counts IN FOV vs. counts OUT FOV. Each point represents an
observation: MOS1 black crosses, MOS2 red diamonds.}  
\label{adeluca-F_fig:cts}
\end{figure}
\\ The main results are contained in Figures \ref{adeluca-F_fig:cts} and 
\ref{adeluca-F_fig:ratio}. 
Figure \ref{adeluca-F_fig:cts} shows that a good correlation exists between 
counts 
IN FOV and OUT FOV; however some scatter is present, possibly associated with
anomalous counts IN FOV. Figure \ref{adeluca-F_fig:ratio} shows that the ratio 
(counts IN FOV)/(counts 
OUT FOV) strongly correlates with the count rate IN FOV: the cluster of points
in the lower left part of the graph represent the correct behaviour, while
the correlation of the scattering points with the count rate IN FOV is a 
possible demonstration of the presence of some SP contamination.
\\ These results can be used in two ways. First, we select the observations
having a stable ratio (IN FOV)/(OUT FOV), within a maximum scatter of 15\%,
rejecting pathological cases. Second, if we suppose that the contaminating 
component has a spectral shape similar to the quiescent NXB (from closed 
observations), the value of the 
ratio (IN FOV)/(OUT FOV) can be used to try a renormalization of the quiescent 
background spectrum  to be subtracted, in order to 
have $${\left( {  \frac {IN}{OUT}  } \right)}_{open} \hskip 0.5in = \hskip 
0.5in {\left( { \frac {IN}{OUT}} \right)}_{closed}$$

\section{Data acquisition, reduction and analysis}
\label{adeluca-F_sec:data}

This work is based on a set of Commissioning, Performance Verification and GT 
observations selected among
high galactic latitude ({$\mid b \mid > 27^{\circ}$}) pointings in order to
avoid contamination by our galaxy; we discarded Magellanic Clouds pointings and
observations of very bright sources. The closed observations collected during 
the very early orbits ($\le$ 40) were rejected, possibly being not fully 
reliable. 
\\We used the XMM-Newton Science Analysis System (XMM -SAS) v.5.0 to perform the 
standard processing of the raw event lists. The linearized event lists were then
cleaned from hot pixels and bad columns using an ad-hoc developed procedure 
which uses cosmic ray IRAF tasks to localize the pixels to be rejected in each 
CCD and XMM-SAS task $evselect$ to remove them. Next step was GTI filtering
to avoid flaring NXB intervals: we set a threshold of 0.29 cts/s in the energy
range 8$\div$12 keV, rejecting also the time bins having 0 counts.
The event files from closed observations were merged in a single list for each
camera.
\\ For each observation a geometric mask was created and applied to reject 
an eventually present bright central source. The diagnostic described in 
\ref{adeluca-F_sec:sub} was then applied to verify the level of SP 
contamination. For each selected observation, we defined the best area to
extract the CXB spectrum by a simple Signal to Noise ratio optimization (we
remember that the signal of CXB is vignetted, while the instrumental NXB 
is not, so the S/N decreases with increasing off-axis angles). The spectrum
were then extracted using an ad-hoc developed task which corrects for the 
telescope vignetting on an event-by-event logic, using the most recent 
vignetting function determinations and accounting for the azimuthal modulation 
induced by the RGA.
The same routine is used to extract the NXB spectrum from the closed event list;
 the region of extraction 
is chosen in order to coincide in detector coordinates to the one used 
for the corresponding observation of the sky.  
\\ For each observation a tentative ``pure'' CXB spectrum is obtained with a 
simple subtraction of the NXB spectrum. A second spectrum is obtained after a 
 renormalization of the NXB spectrum according to the prescription 
described in sect. \ref{adeluca-F_sec:sub}.

\section{Spectral analysis and results}
\label{adeluca-F_sec:spec}

\begin{figure}[ht]
  \begin{center}
    \epsfig{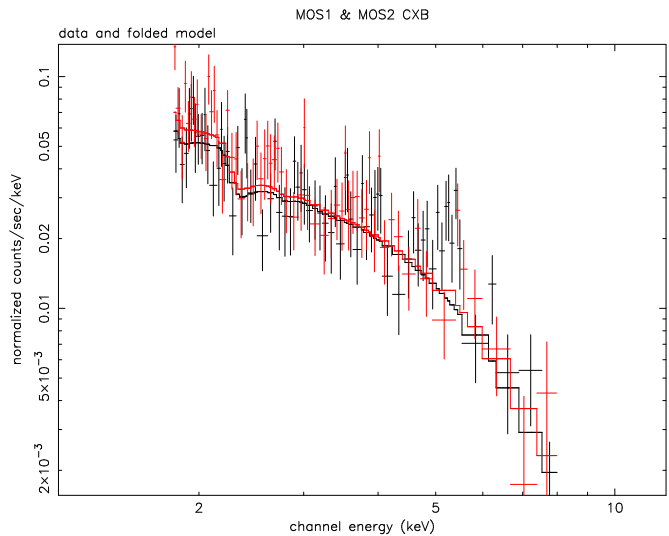}
  \end{center}
\caption{Spectrum of CXB for MOS1 (black) and MOS2 (red); Lockman Hole 
pointing; thick filter; $\approx30$ ks of exposure per camera. The best 
fit power law model is overplotted.}  
\label{adeluca-F_fig:lock}
\end{figure}

The selected observations were 13 for MOS1 for a total good exposure time of 
200 ks and 14 for MOS2 (210 ks).
The CXB spectrum obtained for each observation was fitted using an absorbed 
power law model in the energy band 1.8-8 keV, setting N$_H$ to $4 \times 
10^{20}$ cm$^2$; the normalization was computed at 4 keV. We computed the mean
values of the CXB spectrum slope and intensity for each camera by a straight
$\chi^{2}$ minimization. Figure \ref{adeluca-F_fig:lock} shows the CXB 
spectrum extracted from an observation of the Lockman Hole field as seen by 
MOS1 (black) and MOS2 (red); exposure time is $\approx$ 30 ks per camera, 
the best fit power law model is superposed.
The same analysis was performed separately using the raw NXB spectrum and 
the renormalized one. The results are shown in Table 1.
\begin{table}[ht]
  \label{adeluca-F_tab:tab}
  \begin{center}
\begin{tabular}{|l|c|c|} 
\hline \hline
\multicolumn{3}{|c|}{Non renormalized case} \\
\hline 
 & Photon index & Intensity \\  \hline
 MOS1 & 1.38$\pm$0.10 & 1.27$\pm$0.04  \\  \hline
 MOS2 & 1.44$\pm$0.09 & 1.29$\pm$0.05  \\  \hline 
\multicolumn{3}{|c|}{Renormalized case} \\
 \hline 
 & Photon index & Intensity \\  \hline
 MOS1 & 1.52$\pm$0.11 & 1.01$\pm$0.04  \\  \hline
 MOS2 & 1.62$\pm$0.10 & 1.00$\pm$0.05  \\  \hline \hline
  \end{tabular}
  \end{center}
\caption{Results of the spectral analysis. The parameters for the two cameras
have been computed  with a $\chi^{2}$ minimization on the best fit values 
obtained for each 
observation. The intensity (@ 4 keV) is in units of 1.58 photons s$^{-1}$ 
cm$^{-2}$ keV$^{-1}$ 
sr$^{-1}$, which corresponds (for a photon index equal to 1.4) to 11 photons 
s$^{-1}$ cm$^{-2}$ keV$^{-1}$ 
 sr$^{-1}$ at 1 keV, an average value from previous measurements.}
\end{table}
The {\em non-renormalized} case yields a correct value for the photon index, 
while the normalization is too high by $\approx$ 30\%; on the contrary, the  
{\em renormalized} case yields a lower value for the normalization, but the  
photon index is too hard by $\approx$ 15\%. The 
overall shape of the CXB spectrum has thus been correctly determined, being  
virtually coincident to previous determinations 
(see sect. \ref{adeluca-F_sec:intro}); 
this means that the NXB spectrum has been properly characterized. 
The simple renormalization of NXB spectrum reduces by 30\% the CXB intensity,
but the results suggest that a more sophisticated procedure may be required.
 
\section{Conclusions}
\label{adeluca-F_sec:concl}

Our study of the Cosmic X-ray Background is currently in progress.
We obtained a characterization of the different NXB components
for the MOS cameras. We showed that the standard GTI filtering to remove 
the flaring component possibly 
leaves a low-level SP NXB unrejected, and that the correlation of counts 
(IN FOV)/(OUT FOV) can provide a good diagnostic to identify and evaluate this
effect, allowing to reject pathological cases.  
We tried to correct for SP contamination by means of a simple renormalization 
of the quiescent NXB spectrum.
The preliminary results show that the CXB spectral shape is correct, but
the intensity is too high by $\approx$ 30\%. The renormalization approach 
is encouraging, but will require further study.

\begin{acknowledgements}
We would like to thank Dave Lumb for useful discussions. 
We are also grateful to all the members of the EPIC team for their work and for 
providing useful information and support.

\end{acknowledgements}


\begin{thebibliography}{}

\bibitem[\protect\astroncite{Barcons et~al.}{2000}]{adeluca-F:ba00}
Barcons, X., Mateos, S. and Ceballos, M. T., 2000, MNRAS 316, L13

\bibitem[\protect\astroncite{Gendreau et~al.}{1995}]{adeluca-F:ge95}
Gendreau, K., Mushotzky, R., Fabian, A.C. et al., 1995, PASJ 47, L5 

\bibitem[\protect\astroncite{Georgantopulos et~al.}{1996}]{adeluca-F:ge96}
Georgantopulos, I., Stewart, G., Shanks, T. et al., 1996, MNRAS 280, 276

\bibitem[\protect\astroncite{Giacconi et~al.}{1962}]{adeluca-F:gi62}
Giacconi, R., Gursky, H., Paolini, F. and Rossi, B., 1962, Phys. Rev. Lett. 9, 439

\bibitem[\protect\astroncite{Marshall et~al.}{1980}]{adeluca-F:ma80}
Marshall, F.E., Boldt, E.A., Holt, S.S. et al., 1980, ApJ 235, 4 

\bibitem[\protect\astroncite{Miyaji et~al.}{1998}]{adeluca-F:mi98}
Miyaji, T., Ishisaki, Y., Ogasaka, Y. et al., 1998, A\&A 334, L13 

\bibitem[\protect\astroncite{Pompilio}{2000}]{adeluca-F:po00}
Pompilio, F., La Franca, F. and Matt, G., 2000, A\&A 353, 440

\bibitem[\protect\astroncite{Str\"uder et~al.}{2001}]{adeluca-F:s01}
Str\"uder, L., Briel, U., Dennerl, K. et al., 2001, A\&A,365, L18

\bibitem[\protect\astroncite{Turner et~al.}{2001}]{adeluca-F:tu01}
Turner, M.J.L., Abbey, A., Arnaud, M. et al., 2001, A\&A 365, L27

\bibitem[\protect\astroncite{Vecchi et~al.}{1999}]{adeluca-F:ve99}
Vecchi, A., Molendi, S., Guainazzi, M., Fiore, F. and Parmar, A.N., 1999, A\&A 349, L73

\bibitem[\protect\astroncite{}{}]{adeluca-F:}


\end{thebibliography}
\end{document}